# *In situ* measurement of electrical resistivity evolution during dynamic compression of copper

K. Yan[1,2,3, *], Zhaokun Wang[4, *], Haijuan Mei[5], Wanting Sun[6, *]

[1] Department of Materials Science and Engineering, Southern University of Science and Technology, Shenzhen, Guangdong 518055, China

[2] Department of Mechanical and Aerospace Engineering, The Hong Kong University of Science and Technology, Hong Kong, China

[3] School of Aeronautics, Northwestern Polytechnical University, Xi'an, Shaanxi 710072, China

[4] Department of Mechanical Engineering, The Hong Kong Polytechnic University, Kowloon, Hong Kong, China

[5] Guangdong Provincial Key Laboratory of Electronic Functional Materials and Devices, Huizhou University, Huizhou, Guangdong 516007, China

[6] School of Engineering, Lancaster University, Lancaster, LA1 4YW, UK

*kyanaa@connect.ust.hk (Y. Kai); zhaokun.wang@connect.polyu.hk (Z. Wang); w.sun10@lancaster.ac.uk (W. Sun).

**Abstract:**

We report a novel experimental methodology for *in situ* measurement of electrical resistivity changes in T2 copper during dynamic compression utilizing a split Hopkinson pressure bar. The effects of adiabatic temperature rise and specimen shape deformation on the resistance were carefully accounted, which allowed one to isolate the contribution of microstructure changes such as dislocation evolution, defect generation, and lattice distortion. The latter allows for a real-time relationship between strain and electrical resistivity to be tracked. The experimental findings are also supplemented by molecular dynamics simulations that provide details about the process of microstructure evolution under dynamic loading. Up to now, very few *in situ* measurements has been carried out for changes in electrical resistivity during dynamic deformation, thus establishing a direct link for resistivity-strain which has important implications toward the understanding of plastic deformation and industrial application guidance.

1. **Introduction**

The relationship between mechanical deformation and electrical resistivity is a long-standing topic of interest in materials science, with significant implications for understanding the underlying mechanisms of plastic deformation and developing novel materials with tailored properties [1-3]. Electrical resistivity, a fundamental property of materials, is highly sensitive to changes in microstructure, making it an attractive probe for monitoring deformation-induced microstructural evolution [4-6]. However, the complex interplay between mechanical deformation, temperature, and microstructural changes has hindered the development of a comprehensive understanding of the resistivity-strain relationship, particularly under dynamic loading conditions [2].

High-strain-rate deformation, characterized by rapid loading rates and large deformations, is a critical regime in many industrial applications, including metal forming, machining, and impact resistance [7-10]. In these scenarios, the material's microstructure undergoes significant changes, including dislocation evolution, defect generation, and lattice distortion, which can profoundly impact its electrical resistivity [11-13]. Despite its importance, the real-time measurement of electrical resistivity during dynamic deformation has remained a significant challenge due to the difficulties in isolating the contributions from microstructural changes, adiabatic temperature rise, and specimen shape deformation.

Previous studies have employed various techniques to investigate the relationship between electrical resistivity and deformation, including four-point probe measurements, eddy current testing, and X-ray diffraction [2, 14]. However, these methods often require post-mortem analysis or are limited to quasi-static loading conditions, making it difficult to capture the dynamic evolution of microstructure and resistivity. The development of *in situ* measurement techniques has been hindered by the need to decouple the effects of temperature, deformation, and microstructural changes on electrical resistivity.

Recent advances in experimental techniques, such as the split Hopkinson pressure bar (SHPB), have enabled the study of dynamic deformation under controlled conditions [15-17]. However, the integration of electrical resistivity measurements with SHPB has remained a significant challenge. Although there are some previous study related to this topic [18-21], the lack of a direct link between strain and electrical resistivity under dynamic loading has limited our understanding of the underlying mechanisms of plastic deformation and hindered the development of materials with optimized properties for high-strain-rate applications.

In this work, we report a novel experimental methodology for *in situ* measurement of electrical resistivity changes in T2 copper during dynamic compression using a SHPB. By employing a 1/4 Wheatstone bridge setup, we accurately isolate the contributions from microstructural changes, such as dislocation evolution, defect generation, and lattice distortion, from the effects of adiabatic temperature rise and specimen shape deformation. This approach enables the real-time measurement of electrical resistivity during dynamic deformation, providing new insights into the evolution of microstructure under high-strain-rate loading conditions. The experimental findings are supplemented by molecular dynamics simulations, which provide a detailed understanding of the underlying mechanisms of plastic deformation. Our study establishes a direct link between strain and electrical resistivity, with significant implications for the understanding of plastic deformation and the development of materials with optimized properties for industrial applications.

**2. Materials and Methods**

**2.1. Materials**

Commercial T2 copper were selected for this study. This material was chosen because of its distinct mechanical properties and widespread industrial applications [22-24]. The samples were fabricated via vacuum arc melting and then machined into rectangular shapes with dimensions of 6×6×9 mm before impact loading.

**2.2. Experiment setup**

A SHPB apparatus was used to apply dynamic loading to the samples. The SHPB apparatus consisted of an incident bar, a transmission bar, and a striker bar, which were all made of high-strength steel (Fig. 1a). To operate as a wave shaper, a tiny layer of copper tape was applied to the incident bar's impact end. The transmission bar and the incident bar, which are made of high-strength steel with a diameter of 12 mm and a length of 1.2 m, are positioned between the samples. The striker bar, 150 mm in length, was propelled by compressed gas to impact the incident bar, generating stress waves that were transmitted through the sample.

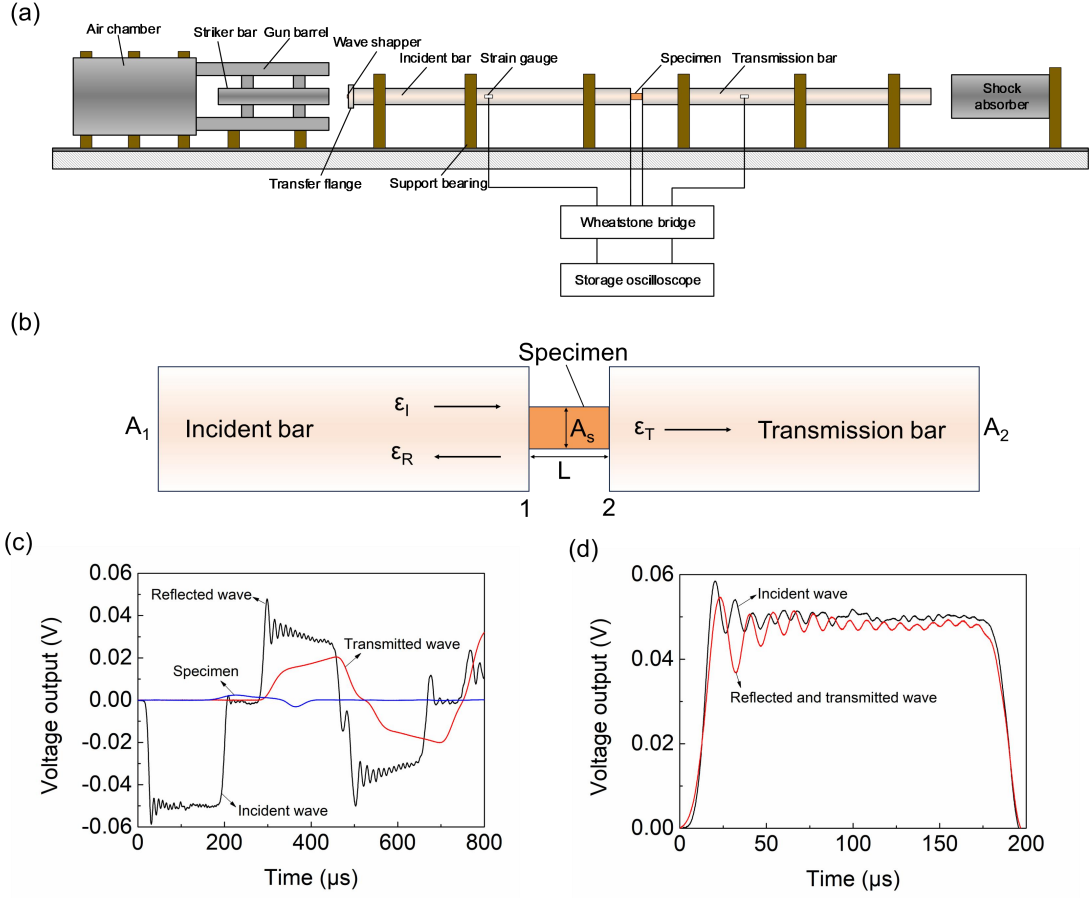

**Fig. 1.** (a) Schematic diagram showing the apparatus for the SHPB experiment. (b) Theoretical model of SHPB test. (c) Voltage-time curves of the wheatstone bridge for the specimen, and the strain gauges glued to the incident, transmission bars, and. (d) Stress balance verification of the specimen.

Fig. 1b shows the theoretical model of the SHPB test. The strain rate, strain, and stress of the tested sample may be computed via one-dimensional elastic stress wave theory as follows [25]:

$$\dot{\varepsilon}(t) = \frac{2C_0}{l_s}\varepsilon_r \quad (1)$$

$$\varepsilon(t) = \frac{2C_0}{l_s}\int_0^t \varepsilon_r dt \quad (2)$$

$$\sigma(t) = -E\frac{A}{A_s}\varepsilon_t \quad (3)$$

where $\varepsilon_t$ and $\varepsilon_r$ are the transmitted and reflected pulse strains measured by the strain gauges glued onto the transmission and incident bars, respectively. $C_0$ and $E$ denote the longitudinal elastic wave velocity and Young's modulus, respectively. $A$ denotes the cross-sectional area of the incident and transmission bars. $A_s$ and $l_s$ are the cross-sectional area and length of the sample, respectively. The detailed deduction process is attached in the appendix. It should be noted that the stress and strain in compression state was set as positive in the convenience of data analysis.

**2.3. Measurement of the electrical resistivity**

*In situ* monitoring of electrical resistivity can be realized through a quarter Wheatstone bridge circuit [26]. Fig. 1c shows the obtained voltage–time curves of the wheatstone bridge for the incident, transmission bars, and specimen during dynamic loading. Fig. 1d shows the stress balance verification during the SHPB experiments. The force is balanced on both sides of the sample, as shown by the similarity in the voltage–time curves of the incident wave and the superposition of the reflected and transmitted waves, which are very close to the superposition of the reflected and transmitted waves.

While Wheastone bridge method is widely employed to get the change of resistance with time in dynamic material testing, it is important to note that the contact resistance between the probes and the specimen can contribute to the overall measured resistance. In order to solve this problem, we calibrated the measure resistance of material with the value measured with theoretical value provided by the manufacturer. The established correlation between resistance and electrical resistivity forms the basis of this approach, the resistivity is governed by the equation [27]:

$$R = \rho \frac{L}{A} \qquad (4)$$

where R is resistance, ρ is electric resistivity, L is specimen length, and A is cross-sectional area. While prior studies have demonstrated this relationship in dynamic experiments are often influenced by complex factors, including adiabatic heating and geometric alterations, which can obscure the underlying mechanisms. With one dimensional stress wave theory, and based on the volume invariant hypothesis, we can get the change of L and A, then the real time change of ρ can be obtained.

**2.3. Simulation setup**

Molecular dynamics (MD) simulations of uniaxially compressed single crystal aluminum were performed using the LAMMPS software. An FCC lattice (with 100 orientation) structure, at a lattice parameter of 3.615 Å, was used to initialize the system. The size of the simulation box was 45×30×30 units. In aluminum, an MEAM potential model was used for describing interatomic interactions [28]. The system was equilibrated during 20,000 steps by 1 fs time step at 300 K. Simultaneously, in an NPT ensemble, we applied a strain rate of $1.0 \times 10^{10}$ s$^{-1}$ along the *x*-direction with zero pressure in the *y*- and *z*-directions. The data for the stress-strain and atomic configurations were written at intervals, and the total simulation time of the present study was 20 ps.

**3. Experimental results**

The change of true stress, resistance, electrical resistivity, true strain, and loading velocity at different loading time is illustrated in Fig. 2. It is seen that the stress value increases linearly from zero to about 310 MPa followed by strain hardening to about 423 MPa, and finally decrease to zero during unloading process. The true strain also shows similar trend, increase from zero to about 41% during loading, only with the maximum value a bit carry-forward comparing with true stress. This should due to the loading velocity is still positive during 170-190 μs, which can be seen from the loading velocity curve.

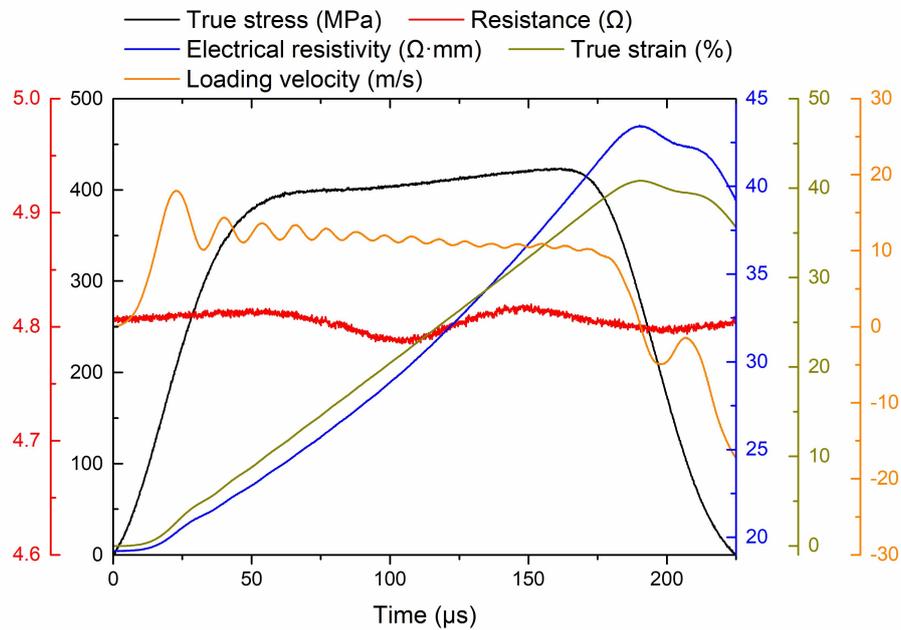

**Fig. 2.** Change of true stress, resistance, electrical resistivity, true strain, and loading velocity at different loading time.

The electrical resistivity is found to increase from about 19.2 ± 0.2 Ω·mm before loading to about 43.4 ± 0.4 Ω·mm at 190 μs, and decrease after unloading. While the change of resistance has no obvious trend with time during the dynamic loading process, which should be counteracted by the increase of electrical resistivity and decrease of length of the specimen, and increase of the cross-sectional area of the specimen.

The change of true stress, resistance, electrical resistivity, true strain, and loading velocity at different loading time is illustrated in Fig. 3. It is seen that the stress value increases almost linearly to about 10% strain, followed by strain hardening, and finally decrease during unloading process.

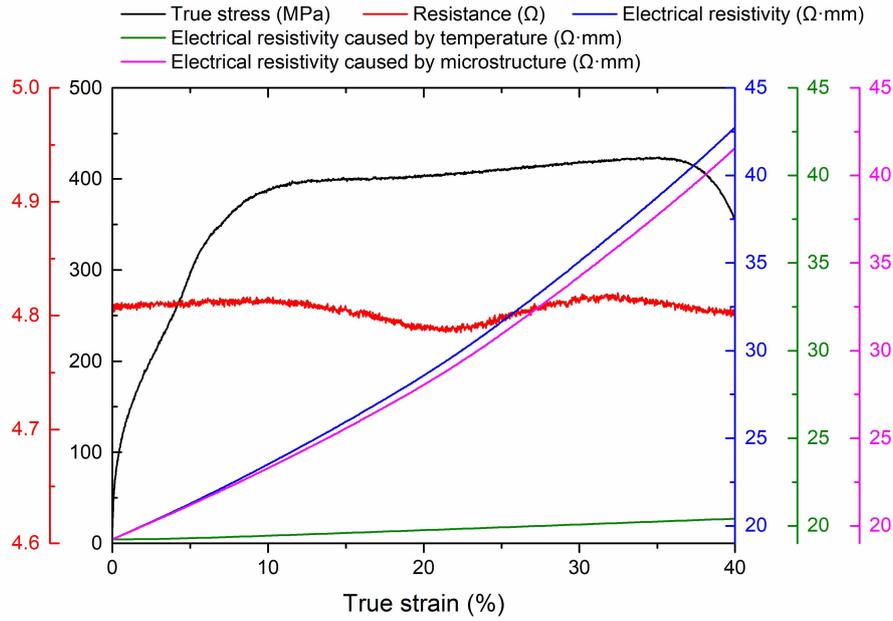

**Fig. 3.** Change of true stress, resistance, electrical resitivity, true strain, and loading velocity at different loading time.

The change of resistance also has no obvious trend with time during the dynamic loading process. The electrical resistivity is found to increase from about 19.2 ± 0.2 Ω·mm before loading to about 43 ± 0.4 Ω·mm at 40% strain. The increase of electric resistivity can be devided into two parts, the electrical resistivity caused by temperature rise and microstructure. The electrical resistivity caused by temperature rise can be calculated by combining the relationship between electrical resistivity and temperature, and the Taylor-Quinney coefficient. It is seen that the change of electrical resistivity caused by temperature is very small, only accounts for ~4% for the total electrical resistivity change. This means the change of electrical resistivity is mainly caused by the microstructure change (such as dislocation evolution, defect generation, and lattice distortion, etc.).

4. **Discussion**

In this study, we developed a new method to measure the electrical resistivity of copper in real-time during dynamic compression using a split Hopkinson pressure bar (SHPB) and a 1/4 Wheatstone bridge setup. This technique allowed us to track how electrical resistivity evolves under high-strain conditions and establish a direct connection between strain and resistivity changes.

Our experiments showed that electrical resistivity significantly increases as the copper undergoes plastic deformation. This rise in resistivity is mainly driven by the generation and multiplication of dislocations, which disrupt the flow of electrons [29]. As more dislocations form, electron scattering increases, causing a spike in resistivity [30]. This is consistent with the principle that defects like dislocations, vacancies, and grain boundaries interfere with electron movement in metals [31].

In copper, the relationship between dislocation density and resistivity can be explained using Matthiessen's rule, which says total resistivity is the sum of the resistivity from the perfect crystal structure and that from scattering sources, like dislocations [32]. Under dynamic loading, dislocations multiply rapidly, causing a noticeable rise in resistivity [33, 34], which aligns with our experimental results showing a sharp increase during plastic deformation.

While dislocations play the main role in resistivity changes, it's important to consider the types of dislocations and how they interact. In our experiments, both screw and edge dislocations likely contribute to deformation. Screw dislocations can move along various planes, significantly impacting the plastic behavior. Edge dislocations likely contribute to lattice distortion and the increase in dislocation density (Fig. 4a-c).

Other defects, like vacancies and stacking faults, also play a role. The high strain rates in our experiments create extreme conditions that lead to non-equilibrium defects, which further hinder electron flow and increase resistivity. Our molecular dynamics (MD) simulations confirmed the presence of these defects, especially vacancies, which act as additional electron scattering centers (Fig. 4d-f).

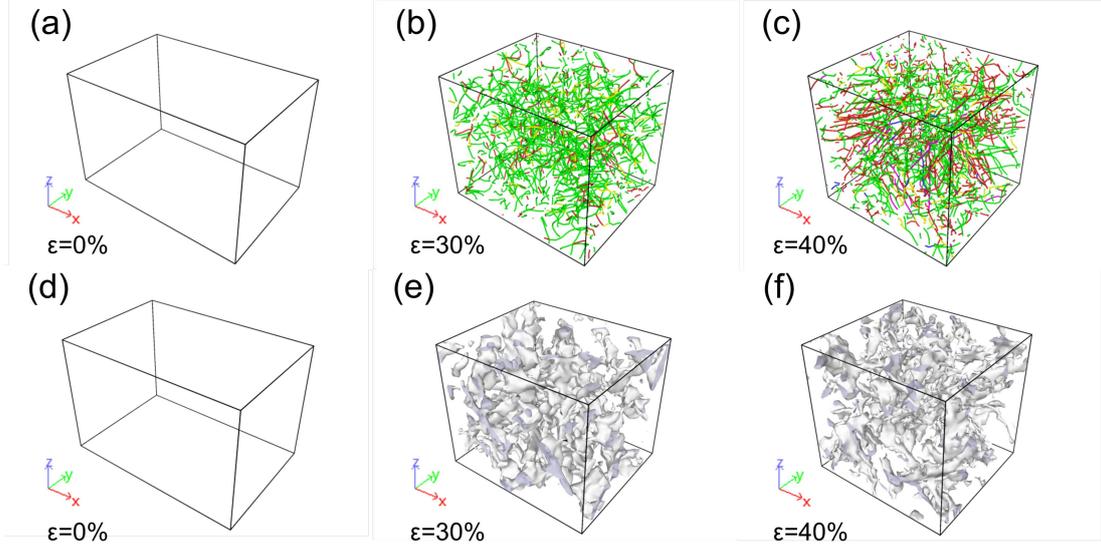

**Fig. 4.** Evolution of dislocations (a-c) and defects (d-f) during dynamic loading of single crystal aluminum under the loading rate of $1\times10^{10}$ s$^{-1}$ at different loading time.

One of the key outcomes of our study is the quantitative link between strain and resistivity. As seen in the experimental data, resistivity increases almost linearly with strain up to a point, after which the rate of increase slows. This suggests that at low strains, dislocation multiplication dominates, but at higher strains, further dislocation movement is limited by interactions like pile-ups.

We describe this relationship with a simple model:

$$\Delta\rho = K\cdot\varepsilon^n \quad (5)$$

where $K$ is a material constant and $n$ is a strain-hardening exponent. For copper, $n$ can be chosen as 1 during the initial phase, showing a near-linear relationship, which becomes less linear at higher strains, consistent with typical strain-hardening behavior in copper. Then we have:

$$\Delta\rho = K\cdot\varepsilon \quad (6)$$

The value *K* is calculated as 52.1 Ω·mm for the T2 copper in this work. So, the linear relationship between strain and electrical resistivity is built.

Adiabatic heating, caused by rapid compression, could also affect copper's resistivity. However, we carefully measured and accounted for temperature changes using thermocouples and infrared thermography to ensure the resistivity changes we observed were primarily due to microstructural changes, not thermal effects.

Similarly, we accounted for changes in the specimen's shape (like cross-sectional area) during deformation by taking real-time measurements. These corrections ensured that the resistivity changes we report reflect the intrinsic microstructural changes happening under dynamic loading.

Molecular dynamics simulations provided additional insights into the microstructural changes driving the observed resistivity increase. The simulations showed significant lattice distortion and defect generation at the atomic level during compression. We observed dislocation nucleation and motion at this scale, directly linking macroscopic strain with the microstructural evolution that impacts resistivity.

The simulations also revealed vacancy clusters and other point defects that contribute to electron scattering, further increasing resistivity. The strong agreement between our simulation and experimental results confirms that dislocation movement and defect generation are the primary factors behind the resistivity changes.

The method we developed for measuring resistivity during dynamic compression has broad implications for understanding how metals behave under extreme conditions. Although our study focused on copper, this approach can be applied to other conductive materials, such as aluminum, steel, and various alloys, where the interaction between dislocations and resistivity is important.

Future research could explore using this technique with different metals and alloys, particularly those with more complex microstructures like dual-phase steels or shape memory alloys. Additionally, applying this approach to other loading conditions, such as tension or shear, would provide further insights into the strain-resistivity relationship observed in this study.

**5. Conclusions**

In the present work, a novel experimental method for *in situ* measurement of electrical resistivity changes in T2 copper during dynamic compressions is developed and demonstrated. Our approach adopted the use of a 1/4 Wheatstone bridge setup that accurately isolates contributions from microstructure changes such as dislocation density, defect evolution, and lattice distortion from the effects of adiabatic temperature rise and deformation of specimen shape to changes in electrical resistivity. Strain-resistivity relationships could be studied, yielding new insight into the real-time evolution of the microstructure under dynamic loading conditions. The MD simulation results further validated the experimental findings to detail the underlying mechanisms of plastic deformation. This study is an important step toward the understanding of the correlation between electrical resistivity and microstructure evolution during high-strain-rate deformation with possible applications in material design and industrial processing where real-time monitoring of deformation is essential.

**Acknowledgments**

This work was financially supported by the National Natural Science Foundation of China (No. 12302095). The authors would like to thank the experimental assistance of Prof. Tao Suo from Northwestern Polytechnical University.